\documentclass[eqsecnum,superscriptaddress,nofootinbib,preprint,11pt]{revtex4-1}

\usepackage{bm} 
\usepackage{amsmath} 
\usepackage{amssymb} 
\usepackage{tikz} 

\usepackage{pgfplots} 
\pgfplotsset{width=10cm,compat=1.8} 
\usepackage[caption=false]{subfig}

\usepackage{hyperref} 

\newcommand{\abs}[1]{{\left|{#1}\right|}} 
\newcommand{\ket}[1]{\vert{#1}\rangle} 
\newcommand{\bra}[1]{\langle{#1}\vert} 

\newcommand{\secref}[1]{Sec.~\ref{#1}}
\newcommand{\eqnref}[1]{(\ref{#1})}
\newcommand{\figref}[1]{Fig.~\ref{#1}}

\begin{document}

\title
{An elementary rigorous proof of bulk-boundary correspondence in the generalized Su-Schrieffer-Heeger model}


\author{Bo-Hung Chen}
\email{kenny81778189@gmail.com}
\affiliation{Department of Physics, National Taiwan University, Taipei 10617, Taiwan}
\affiliation{Department of Physics, National Taiwan Normal University, Taipei 11677, Taiwan}

\author{Dah-Wei Chiou}
\email{dwchiou@gmail.com}
\affiliation{Department of Physics, National Taiwan Normal University, Taipei 11677, Taiwan}
\affiliation{Department of Physics, National Sun Yat-sen University, Kaohsiung 80424, Taiwan}
\affiliation{Center for Condensed Matter Sciences, National Taiwan University, Taipei 10617, Taiwan}


\begin{abstract}
We generalize the Su-Schrieffer-Heeger (SSH) model with the inclusion of arbitrary long-range hopping amplitudes, providing a simple framework to investigate \emph{arbitrary} adiabatic deformations that preserve the chiral symmetry upon the bulk energy bands with any \emph{arbitrary} winding numbers. Using only elementary techniques of solving linear difference equations and applying Cauchy's integral formula, we obtain a mathematically rigorous and physically transparent proof of the bulk-boundary correspondence for the generalized SSH model. The multiplicity of robust zero-energy edge modes is shown to be identical to the winding number. On the other hand, nonzero-energy edge modes, if any, are shown to be unstable under adiabatic deformations and not related to the topological invariant. Furthermore, under deformations of small spatial disorder, the zero-energy edge modes remain robust.
\end{abstract}


\maketitle

\section{Introduction}
One of the most significant features of topological insulators and quantum Hall systems is the \emph{bulk-boundary correspondence}, which posits that the multiplicities of edge modes on the boundary are characterized by the topological invariants of the bulk energy bands. It has been affirmed in many different experiments and numerical simulations. (See \cite{Hasan:2010xy,Qi:2011} for reviews.) Meanwhile, since Laughlin proposed an explanation for the integer quantum Hall effect in 1981 \cite{Laughlin:1981jd}, many theoretical arguments for the bulk-boundary correspondence have been developed from different aspects with various degrees of rigor (see e.g.\ \cite{Hatsugai:1993ywa,Hatsugai:2009ywa,Essin:2011ywa,Graf:2013ywa,Rudner:2013ywa,Cano:2014ywa} and more references in \cite{Hasan:2010xy,Qi:2011}).

A mathematically rigorous proof of the bulk-boundary correspondence for topological insulators is rather difficult, even for a specific model. The major difficulty lies in the fact that the notions of edge modes and topological invariants are anchored to two different and conflicting settings. Rigorously speaking, only in the explicit presence of boundaries can one make sense of edge modes. On the other hand, the topological invariants are defined on the bulk energy bands, which make sense only if the system is without explicit boundaries and thus respects the lattice translational symmetry --- i.e., either the system is infinite or the system is finite with the periodic (Born-von Karman) boundary condition imposed. As one cannot maintain both notions in a single setting, it is rather challenging to rigorously prove the robustness of the bulk-boundary correspondence. Many advanced mathematical tools have been employed to overcome the difficulty, and nowadays the bulk-boundary correspondence is perhaps best encoded in terms of the $K$-theory (see \cite{Prodan:book} for a review).

The advanced approaches such as the $K$-theory, although rigorous, powerful, and broad in scope, involve heavy technicalities and are often not very transparent about the underlying mechanism. In this paper, we aim to offer a rigorous yet elementary proof of the bulk-boundary correspondence in the generalized Su-Schrieffer-Heeger (SSH) model without invoking any advanced techniques. The SSH model \cite{Su:1979} provides a simple yet paradigmatic example of a one-dimensional system that exhibits nontrivial topological features \cite{Heeger:1988zz,Jackiw:1976,Ryu:2010zza}. (Also see \cite{Asboth:book} for a detailed review.) The SSH model is generalized with the inclusion of long-range hopping amplitudes, making it possible to study \emph{arbitrary} adiabatic deformations upon the bulk energy bands with \emph{arbitrary} winding numbers.

Thanks to the simplicity of the generalized SSH model, we obtain a detailed description and a rigorous proof of the bulk-boundary correspondence using only basic mathematical techniques of solving linear difference equations and applying Cauchy's integral formula. Our elementary approach offers a transparent and instructive perspective on the mechanism of the bulk-boundary correspondence. (The efforts in the similar spirit to give elementary explanations of the bulk-boundary correspondence can also be found in \cite{Pershoguba:2012,Rhim:2017}. For the $K$-theory approach in the one-dimensional case, see Chapter 1 of \cite{Prodan:book}.)

\section{Generalized Su-Schrieffer-Heeger model}
\begin{figure}


\begin{tikzpicture}[scale=0.68]

 \foreach \i in {0,1,2,3,7,8,9,10}{
  \begin{scope}[shift={(2*\i,0)}]
   \draw [blue] (0,0) -- (1,0.7);
   \end{scope}
  }

 \foreach \i in {0,1,2,3,6,7,8,9}{
  \begin{scope}[shift={(2*\i,0)}]
   \draw [blue] (1,0.7) -- (2,0);
   \end{scope}
  }

 \foreach \i in {0,1,2,3,4,7,8,9,10}{
  \begin{scope}[shift={(2*\i,0)}]
   \draw [cyan,fill=cyan] (0,0) circle [radius=0.15];
   \end{scope}
  }

 \foreach \i in {0,1,2,3,6,7,8,9,10}{
  \begin{scope}[shift={(2*\i,0)}]
   \draw [cyan,fill=white] (1,0.7) circle [radius=0.15];
   \end{scope}
  }


 \node [] at (10.5,0.35) {$\cdots\cdots\cdots\cdots$};

 \node [below] at (0,-0.2) {{\scriptsize $\ket{1,A}$}};
 \node [below] at (2,-0.2) {{\scriptsize $\ket{2,A}$}};
 \node [below] at (4,-0.2) {{\scriptsize $\ket{3,A}$}};
 \node [above] at (1,0.9) {{\scriptsize $\ket{1,B}$}};
 \node [above] at (3,0.9) {{\scriptsize $\ket{2,B}$}};
 \node [above] at (5,0.9) {{\scriptsize $\ket{3,B}$}};
 \node [above] at (18.8,0.9) {{\scriptsize $\ket{N\!-\!1,B}$}};
 \node [above] at (21.2,0.9) {{\scriptsize $\ket{N,B}$}};
 \node [below] at (17.8,-0.2) {{\scriptsize $\ket{N\!-\!1,A}$}};
 \node [below] at (20.2,-0.2) {{\scriptsize $\ket{N,A}$}};

\end{tikzpicture}

\caption{The one-dimensional lattice of the SSH model. Each unit cell of the chain consists of two sublattice sites of type $A$ (solid dots) and type $B$ (hollow dots).}
\label{fig:SSH model}
\end{figure}
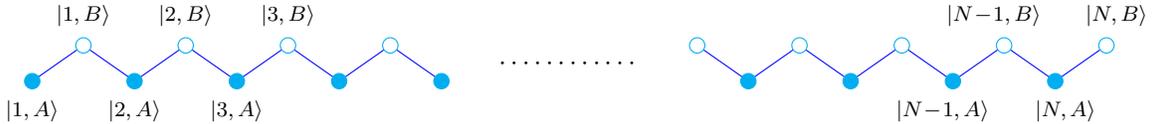

The SSH model \cite{Su:1979} describes spinless fermions hopping on a chain (one-dimensional lattice), where each unit cell hosts two sublattice sites --- one of type $A$ and the other of type $B$ --- as shown in \figref{fig:SSH model}. The hopping amplitudes are ``bipartite'' in the sense that fermions at sublattice $A$ can only hop to sublattice $B$ and \textit{vice versa} (they do not hop from $A$ to $A$ or from $B$ to $B$).

In the standard SSH model, there are two kinds of bipartite hopping amplitudes: intracell hopping within the same cell and intercell hopping to the nearest-neighbor cell.\footnote{We follow closely the lines of Chapter 1 in \cite{Asboth:book}, to which readers are referred for more details of the SSH model.} We generalize the SSH model by including arbitrary ``long-range'' intercell hopping amplitudes that respect the bipartite property.

\subsection{Bulk momentum-space Hamiltonian}\label{sec:momentum-space Hamiltonian}
To begin with, we neglect all boundary effects and study only the physics in the bulk. That is, we either consider an infinite system or impose the periodic (Born-von Karman) boundary condition.
In this idealized setting, the lattice momentum is a good quantum number and the SSH model is described by a single-particle Hamiltonian, which takes the form $\hat{H}_\mathrm{bulk}=\sum_k\hat{H}(k)\ket{k}\bra{k}$ in the bulk momentum space. The bulk momentum-space Hamiltonian is given by
\begin{equation}\label{H(k)}
  \hat{H}(k) := \bra{k}\hat{H}_\mathrm{bulk}\ket{k}
  =\sum_{\alpha,\beta\in\{A,B\}} \bra{k,\alpha}\hat{H}_\mathrm{bulk}\ket{k,\beta} \ket{\alpha}\bra{\beta}.
\end{equation}
We generalize $\hat{H}(k)$ to the generic form
\begin{equation}\label{d(k)}
  \hat{H}(k) = \mathbf{d}(k)\cdot\boldsymbol{\sigma}
  = \left(
      \begin{array}{cc}
        0 & h(k)^* \\
        h(k) & 0 \\
      \end{array}
    \right)
  \equiv h(k)^*\ket{A}\bra{B}+h(k)\ket{B}\bra{A},
\end{equation}
where
\begin{equation}\label{h(k)}
  h(k) \equiv d_x(k)+id_y(k) := \sum_{n=-\infty}^\infty w_n e^{ink},
  \quad w_n\in\mathbb{C}.
\end{equation}
Obviously, the bulk energy spectrum is given by $\epsilon=\pm\abs{\mathbf{d}(k)}$. If $\mathbf{d}(k)=0$ at some point $k$, the energy gap will close at this point and the system is no longer a bulk insulator.

The bulk-boundary correspondence is said to be robust under \emph{adiabatic deformations}, which are defined as any continuous deformations upon the insulating bulk energy spectrum that maintain the important symmetry and keep the bulk energy gap open. The important symmetry for the (generalized) SSH model is the \emph{chiral symmetry} (also known as \emph{sublattice symmetry}), which dictates that the $z$-component of $\mathbf{d}(k)$ remains zero. The winding number of the bulk energy spectrum is invariant under adiabatic deformations.

As the Fourier series \eqnref{h(k)} can represent any generic function mapping from $[-\pi,\pi]$ to $\mathbb{C}$ with $h(k+2\pi)=h(k)$, the form of \eqnref{h(k)} provides a starting point to study any \emph{arbitrary} adiabatic deformations upon the bulk energy spectrum with any \emph{arbitrary} winding numbers.\footnote{Our goal is to obtain a \emph{mathematically rigorous} proof of the bulk-boundary correspondence. Therefore, we have to take into consideration \emph{all} arbitrary adiabatic deformations, even if the corresponding $\hat{H}(k)$ with an arbitrary $h(k)$ is purely fictitious and cannot be realized in a realistic system.} The standard SSH model corresponds to $w_0=v\in\mathbb{R}$, $w_1=w\in\mathbb{R}$, and $w_n=0$ for $n\neq0,1$.

If we deal with a finite system with $N$ unit cells, $k$ takes discrete values $k\in\{\delta_k,2\delta_k,\dots,N\delta_k\}$ with $\delta_k=2\pi/N$, and it is only an approximation to treat $h(k)$ as a continuous map when $N$ is large but finite. To make this approximation sensible, the map $h(k)$ has to be ``smooth'' enough, or more precisely, $\abs{h'(k)/h(k)}\ll1/\delta_k$. This requires $\sum_{n=-\infty}^\infty$ to be truncated to $\sum_{n=-n_l}^{n_r}$ with two integers $n_l,n_r\ll N$.

\subsection{Bulk real-space Hamiltonian}
To study the physics in the bulk for a finite system while neglecting the physics on the boundary, we impose the periodic boundary condition: i.e., $\ket{m+N,A}\equiv\ket{m,A}$ and $\ket{m+N,B}\equiv\ket{m,B}$. As the periodic boundary condition respects the lattice translational invariance, Bloch's theorem applies. The Bloch's theorem allows us to introduce the plane wave basis states
\begin{equation}\label{ket k}
  \ket{k}=\frac{1}{\sqrt{N}}\sum_{m=1}^N e^{imk} \ket{m},
\end{equation}
so that the Bloch eigenstates (labeled by $\epsilon$ and $k$) read as
\begin{equation}
  \ket{\Psi_\epsilon(k)} = \ket{k}\otimes\ket{u_\epsilon(k)},\qquad
  \text{where}\ \ket{u_\epsilon(k)} = a_\epsilon(k)\ket{A}+b_\epsilon(k)\ket{B}.
\end{equation}
The vectors $\ket{u_\epsilon(k)}$ are eigenstates of $\hat{H}(k)$ defined in \eqnref{H(k)}; i.e., $\hat{H}(k)\ket{u_\epsilon(k)}=\epsilon(k)\ket{u_\epsilon(k)}$.

Substituting \eqnref{ket k} into $\hat{H}_\mathrm{bulk}=\sum_k\hat{H}(k)\ket{k}\bra{k}$ with \eqnref{d(k)} and \eqnref{h(k)}, we obtain the bulk real-space Hamiltonian:
\begin{equation}\label{H bulk}
  \hat{H}_\mathrm{bulk}
  = \sum_{m=1}^N \sum_{n=-n_l}^{n_r} w_n^*\ket{m+n,A}\bra{m,B}
  + \sum_{m=1}^N \sum_{n=-n_l}^{n_r} w_n\ket{m,B}\bra{m+n,A}.
\end{equation}

\begin{figure}


\begin{tikzpicture}[scale=0.68]

 \foreach \i in {0,1,2,6,7,8}{
  \begin{scope}[shift={(2*\i,0)}]
   \draw [magenta] (1,0.7) -- (4,0);
   \end{scope}
  }

 \foreach \i in {0,1,2,3,4,7,8,9,10}{
  \begin{scope}[shift={(2*\i,0)}]
   \draw [cyan,fill=cyan] (0,0) circle [radius=0.15];
   \end{scope}
  }

 \foreach \i in {0,1,2,3,6,7,8,9,10}{
  \begin{scope}[shift={(2*\i,0)}]
   \draw [cyan,fill=white] (1,0.7) circle [radius=0.15];
   \end{scope}
  }


 \node [] at (10.5,0.35) {$\cdots\cdots\cdots\cdots$};

 \node [below] at (0,-0.2) {{\scriptsize $\ket{1,A}$}};
 \node [below] at (2,-0.2) {{\scriptsize $\ket{2,A}$}};
 \node [below] at (4,-0.2) {{\scriptsize $\ket{3,A}$}};
 \node [above] at (1,0.9) {{\scriptsize $\ket{1,B}$}};
 \node [above] at (3,0.9) {{\scriptsize $\ket{2,B}$}};
 \node [above] at (5,0.9) {{\scriptsize $\ket{3,B}$}};
 \node [above] at (18.8,0.9) {{\scriptsize $\ket{N\!-\!1,B}$}};
 \node [above] at (21.2,0.9) {{\scriptsize $\ket{N,B}$}};
 \node [below] at (17.8,-0.2) {{\scriptsize $\ket{N\!-\!1,A}$}};
 \node [below] at (20.2,-0.2) {{\scriptsize $\ket{N,A}$}};

\end{tikzpicture}

\caption{Long-range hopping amplitudes. Here, as an example, the amplitudes associated with $w_{n=2}$ and $w^*_{n=2}$ are depicted by the solid lines. Note that, if $w_n=0$ for all $n\in\mathbb{Z}$ except $n=m$, the system is fully dimerized and there are $m$ dangling $A$ ($B$) sites if $m>0$ or $-m$ dangling $B$ ($A$) sites if $m<0$ at the left (right) edge.}
\label{fig:long range amp}
\end{figure}
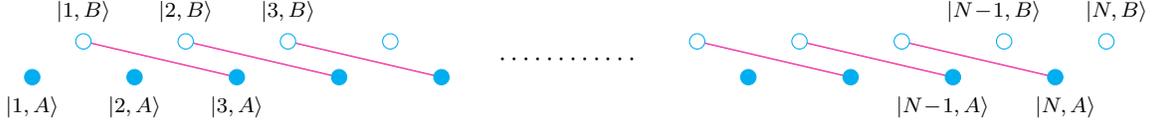

Therefore, the physical meaning of $w_n$ is the hopping amplitude from $A$ in the $(m+n)$-th cell to $B$ in the $m$-th cell; correspondingly, $w_n^*$ is the hopping amplitude from $B$ in the $m$-th cell to $A$ in the $(m+n)$-th cell. (See \figref{fig:long range amp}.) Particularly, $w_0$ and $w_0^*$ are for the intracell hopping. The SSH model is generalized with inclusion of long-range bipartite hopping.

Because the Hamiltonian does not have any terms $\ket{m,A}\bra{m',A}$ or $\ket{m,B}\bra{m',B}$, the generalized SSH model respects the \emph{chiral symmetry} as the standard model does. That is, defining
\begin{equation}
  \hat{\Gamma}:=\sum_{m=1}^N \ket{m,A}\bra{m,A} - \sum_{m=1}^N \ket{m,B}\bra{m,B}
\end{equation}
we have
\begin{equation}\label{chiral symmetry}
  \hat{\Gamma}\hat{H} = -\hat{H}\hat{\Gamma},
\end{equation}
if the Hamiltonian $\hat{H}$ does not contain $\ket{m,A}\bra{m',A}$ or $\ket{m,B}\bra{m',B}$. As a consequence of chiral symmetry, for any eigenstate $\ket{\psi}$ of $\hat{H}$ with energy $\epsilon$, there is a chiral symmetric counterpart $\hat{\Gamma}\ket{\psi}$ with energy $-\epsilon$. If $\epsilon=0$, the corresponding eigenstates $\ket{\psi}$ and $\hat{\Gamma}\ket{\psi}$ are degenerate and can be reshuffled as $(\ket{\psi}\pm\hat{\Gamma}\ket{\psi})/\sqrt{2}$, which have support only in sublattice $A$ and sublattice $B$, respectively, because obviously $(1\pm\Gamma)/2$ are the projection operators that project states into sublattice $A$ and sublattice $B$, respectively. Meanwhile, if an eigenstate has support only at $A$ or at $B$, the eigenvalue must be $\epsilon=0$ because in this case we have $\Gamma\ket{\psi}=\pm\ket{\psi}\propto\ket{\psi}$ and thus $\epsilon=-\epsilon$.

\section{Winding number}
The topological nontriviality of the bulk energy spectrum can be characterized by the \emph{winding number} of $\mathbf{d}(k)$ in \eqnref{d(k)}, as $\mathbf{d}$ is viewed as a map $\mathbf{d}: k\in S^1 \mapsto \mathbf{d}(k)\in\mathbb{R}^2\setminus\{0\}$, where $\mathbf{d}(k)=0$ is excluded to have a bulk energy gap.

The winding number can be expressed as the integral of the complex logarithm function of $h(k)$ (see \cite{Rudin:1976}):
\begin{equation}\label{nu}
  \nu = \frac{1}{2\pi i}\int_{-\pi}^{\pi} dk \frac{d}{dk}\log h(k)
  = \frac{1}{2\pi i}\int_{-\pi}^{\pi} dk
  \frac{\sum_{n=-n_l}^{n_r} i n w_n e^{ink}}{\sum_{n=-n_l}^{n_r} w_n e^{ink}}.
\end{equation}
By rewriting $z=e^{ik}$, $dz=ie^{ik}dk$ and $f(z)=\sum_{n=-n_l}^{n_r} w_n z^n$, the winding number can be recast as a contour integral along the unit circle on the complex plane:
\begin{equation}\label{nu 2}
  \nu = \frac{1}{2\pi i}\oint_{\abs{z}=1} dz \frac{f'(z)}{f(z)}.
\end{equation}

Note that $z^{n_l}f(z)$ is a polynomial with complex coefficients and can be formally factorized as
\begin{equation}\label{factorization}
  z^{n_l}f(z)=\sum_{n=-n_l}^{n_r} w_n z^{n+n_l} = w_{n_r} \prod_j (z-\xi_j)^{\nu_j},
\end{equation}
where $\xi_i$ are the roots of $z^{n_l}f(z)$ and $\nu_i\in\mathbb{N}$ are the corresponding multiplicities. Substituting \eqnref{factorization} for $f(z)$ into \eqnref{nu 2} leads to
\begin{equation}
  \nu = \sum_{j} \frac{1}{2\pi i}\oint_{\abs{z}=1} dz \frac{\nu_j}{z-\xi_j}
  - \frac{1}{2\pi i}\oint_{\abs{z}=1} dz \frac{n_l}{z}.
\end{equation}
Cauchy's integral formula then implies
\begin{equation}\label{key eq 1}
  \nu = -n_l + \sum_{j=1,\dots \atop \abs{\xi_j}<1} \nu_j,
  \qquad \text{where}\ \sum_{n=-n_l}^{n_r} w_n z^{n+n_l} \propto \prod_j (z-\xi_j)^{\nu_j}.
\end{equation}
That is, the winding number is the sum of the multiplicities of those roots of $\sum_{n=-n_l}^{n_r} w_n z^{n+n_l}$ that are located inside the unit circle on the complex plane.\footnote{Note that $\abs{\xi_j}\neq1$ for all $\xi_j$ in \eqnref{factorization}. If $\abs{\xi_j}=1$, we would have $\xi_j=e^{i\theta}$ for some $\theta\in[-\pi,\pi]$ and therefore $h(k=\theta)=f(z=e^{i\theta})=0$, which violates the assumption $h(k)\neq0$.}

Similarly, repeating the above calculation with $z=e^{-ik}$, $dz=-ie^{-ik}dk$ and $f(z)=\sum_{n=-n_l}^{n_r} w_n z^{-n}\equiv\sum_{n=-n_r}^{n_l} w_{-n} z^n$, we obtain a different expression:
\begin{equation}\label{key eq 2}
\nu = n_r - \sum_{j=1,\dots \atop \abs{\xi_j}<1} \nu_j,
\qquad \text{where}\ \sum_{n=-n_r}^{n_l} w_{-n} z^{n+n_r} \propto \prod_j (z-\xi_j)^{\nu_j}.
\end{equation}

Equivalently, the winding number can also be expressed in terms of $h(k)^*$ as
\begin{equation}
  \nu = -\frac{1}{2\pi i}\int_{-\pi}^{\pi} dk \frac{d}{dk}\log h(k)^*
  = -\frac{1}{2\pi i}\int_{-\pi}^{\pi} dk
  \frac{\sum_{n=-n_l}^{n_r} -i n w_n^* e^{-ink}}{\sum_{n=-n_l}^{n_r} w_n^* e^{-ink}}.
\end{equation}
Consequently, we have
\begin{equation}\label{key eq 3}
\nu = -n_l + \sum_{j=1,\dots \atop \abs{\xi_j}<1} \nu_j,
\qquad \text{where}\ \sum_{n=-n_l}^{n_r} w_n^* z^{n+n_l} \propto \prod_j (z-\xi_j)^{\nu_j},
\end{equation}
and
\begin{equation}\label{key eq 4}
\nu = n_r - \sum_{j=1,\dots \atop \abs{\xi_j}<1} \nu_j,
\qquad \text{where}\ \sum_{n=-n_r}^{n_l} w_{-n}^* z^{n+n_r} \propto \prod_j (z-\xi_j)^{\nu_j}.
\end{equation}
Eqs.~\eqnref{key eq 1}, \eqnref{key eq 2}, \eqnref{key eq 3} and \eqnref{key eq 4} are the key identities that will be used to relate the winding number and the multiplicity of the zero-energy edge states.

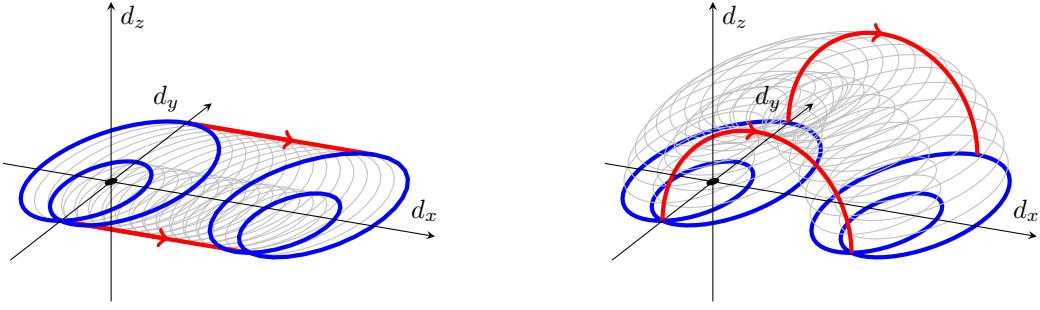
\begin{figure}
  \centering

  \begin{tikzpicture}
  \begin{scope}[shift={(0,0)},scale=1]
  \begin{axis}
    [
    axis lines = center,
    axis on top,
    xlabel={$d_x$},
    ylabel={$d_y$},
    zlabel={$d_z$},
    xtick   = \empty,
    ytick   = \empty,
    ztick   = \empty,
    xmin    = -2,
    ymin    = -2,
    zmin    = -2,
    xmax    =  6,
    ymax    =  2,
    zmax    =  3
    ]
 \foreach \i in {1,2,3,4,5,6} 
  {

   \addplot3[
    domain=0:2*pi,
    samples = 60,
    samples y=0,
    thick,
    gray!50,
  ]
  ({0.5*\i+0.5*sin(deg(x))+sin(deg(2*x))},
  {0.5*cos(deg(x))+cos(deg(2*x))},
  {0});

  }

  \draw[ultra thick,red] (axis cs:0,1.5,0) --(axis cs:3.5,1.5,0);
  \draw[ultra thick,red] (axis cs:0,-1,0) --(axis cs:3.5,-1,0);
  \draw[ultra thick,red,->] (axis cs:0,1.5,0) --(axis cs:2,1.5,0);
  \draw[ultra thick,red,->] (axis cs:0,-1,0) --(axis cs:2,-1,0);
  \draw [fill] (axis cs:0,0,0) circle [radius=0.08];

\addplot3[
    domain=0:2*pi,
    samples = 100,
    samples y=0,
    ultra thick ,
    blue
]
({0.5*sin(deg(x))+1*sin(deg(2*x))},
{0.5*cos(deg(x))+1*cos(deg(2*x))},
{0});

\addplot3[
    domain=0:2*pi,
    samples = 60,
    samples y=0,
    ultra thick ,
    blue
]
({3.5+0.5*sin(deg(x))+1*sin(deg(2*x))},
{0.5*cos(deg(x))+1*cos(deg(2*x))},
{0});
\end{axis}
\end{scope}

\begin{scope}[shift={(8,0)},scale=1]

 \begin{axis}
    [
    axis lines = center,
    axis on top,
    xlabel={$d_x$},
    ylabel={$d_y$},
    zlabel={$d_z$},
    xtick   = \empty,
    ytick   = \empty,
    ztick   = \empty,
    xmin    = -2,
    ymin    = -2,
    zmin    = -2,
    xmax    =  6,
    ymax    =  2,
    zmax    =  3
    ]
    \addplot3[
    domain=0:2*pi,
    samples = 100,
    samples y=0,
    ultra thick,
    blue
    ]
    ({0.5*sin(deg(x))+sin(deg(2*x))},
    {0.5*cos(deg(x))+cos(deg(2*x))},
    {0});

    \addplot3[
    domain=0:2*pi,
    samples = 100,
    samples y=0,
    ultra thick,
    blue
    ]
    ({3.5+0.5*sin(deg(x))+sin(deg(2*x))},
    {0.5*cos(deg(x))+cos(deg(2*x))},
    {0});

    \foreach \i in {1,2,3,4,5,6,7,8} 
    {
    \addplot3[
    domain=0:2*pi,
    samples = 100,
    samples y=0,
    thick,
    gray!50,
    ]
    ({1.75+1.75*cos(20*\i)+0.5*sin(deg(x))+sin(deg(2*x))},
    {0.5*cos(deg(x))+cos(deg(2*x))},
    {1.75*sin(20*\i)});
    }

    \foreach \i in {0,1}{
    \addplot3[
     domain=0:pi,
     samples = 60,
     samples y=0,
     ultra thick,
     red
    ]
    ({1.75+1.75*cos(deg(x))},
    {-1+2.5*\i},
    {1.75*sin(deg(x))});

    }
    \draw[ultra thick,red,->] (axis cs:1.73,-1,1.75) --(axis cs:1.76,-1,1.75);
    \draw[ultra thick,red,->] (axis cs:1.73,1.5,1.75) --(axis cs:1.76,1.5,1.75);
    \draw [fill] (axis cs:0,0,0) circle [radius=0.08];
\end{axis}
\end{scope}
\end{tikzpicture}

\caption{Deformations upon the trajectory of $\mathbf{d}(k)$. \textit{Left}: If the constraint $d_z=0$ is imposed, the trajectory remains on the $d_x$-$d_y$ plane. Under deformations (depicted here as continuous translations), the winding number is unchanged unless the trajectory of $\mathbf{d}(k)$ passes over the origin. \textit{Right}: If $d_z\neq0$ is allowed, the trajectory of a particular winding number can be continuously deformed into a new one of a different winding number without touching the origin.}
\label{fig:non-bipartite}
\end{figure}

What happens if we also include hopping processes that violate the bipartite property? If the hopping amplitudes from $A$ to $A$ or from $B$ to $B$ are allowed, the diagonal entries of the $2\times2$ matrix in \eqnref{d(k)} will no longer be identically zero. Accordingly, $d_z$ is not identically zero and $\mathbf{d}(k)$ should be viewed as a map $\mathbf{d}: k\in S^1 \mapsto \mathbf{d}(k)\in\mathbb{R}^3\setminus\{0\}$ instead of $\mathbb{R}^2\setminus\{0\}$.
As illustrated in \figref{fig:non-bipartite}, the winding number is well defined and unchanged under arbitrary adiabatic deformations if the constraint $d_z=0$ is imposed. However, if $d_z\neq0$ is allowed, $\mathbf{d}(k)$ with a particular winding number can always be continuously deformed into a new configuration with a different winding number without touching the origin no matter how small $\abs{d_z}$ is. That is, the winding number is no longer an invariant under arbitrary adiabatic deformations. As a consequence, we cannot make sense of the bulk-boundary correspondence characterized by the winding number.
If nonbipartite hopping amplitudes are nonzero but remain small enough, the winding number can still be viewed as an invariant in the \emph{approximate} sense that, instead of taking into account \emph{any} arbitrary adiabatic deformations, the adiabatic deformations are assumed to satisfy $\abs{d_z}\ll\abs{d_x+id_y}$. In this sense, the bulk-boundary correspondence remains a good approximation.
On the other hand, if there is some mechanism that renders the chiral symmetry exact, $d_z$ is identically zero and the bulk-boundary correspondence is truly exact.

\section{Exact calculation of zero-energy edge modes}\label{sec:exact calculation}
To study the physics not only for the bulk but also for the boundaries, we should not impose the periodic boundary condition (which is artificial for a finite system). Without the periodic boundary condition, the lattice points close to the boundaries are no longer on the equal footing as those in the bulk. We have to take special care of the modifications upon \eqnref{H bulk} for the left and right ``margins''. As a result, the Hamiltonian of the finite system with $N$ cells is given by
\begin{eqnarray}\label{HN}
  \hat{H}_N
  &=& \sum_{m=1}^N  w_0^*\ket{m,A}\bra{m,B} + \sum_{m=1}^N w_0\ket{m,B}\bra{m,A} \nonumber\\
  && \mbox{} + \sum_{n=1}^{n_r} \sum_{m=1}^{N-n} w_n^*\ket{m+n,A}\bra{m,B}
             + \sum_{n=1}^{n_r} \sum_{m=1}^{N-n} w_n\ket{m,B}\bra{m+n,A} \nonumber\\
  && \mbox{} + \sum_{n=1}^{n_l} \sum_{m=n+1}^{N} w_{-n}^*\ket{m-n,A}\bra{m,B}
             + \sum_{n=1}^{n_l} \sum_{m=n+1}^{N} w_{-n}\ket{m,B}\bra{m-n,A}.
\end{eqnarray}
Note that $\hat{H}_N$ still have the chiral symmetry \eqnref{chiral symmetry}.

The eigenvalue problem of $\hat{H}_N\ket{\psi}=\epsilon\ket{\psi}$ with
\begin{equation}
  \ket{\psi}=\sum_{m=1}^N \left(a_m\ket{m,A}+b_m\ket{m,B}\right),
\end{equation}
reads as
\begin{eqnarray}
  && \sum_{m=1}^N w_0 a_m \ket{m,B}
  + \sum_{n=1}^{n_r} \sum_{m=1}^{N-n} w_n a_{m+n} \ket{m,B}
  + \sum_{n=1}^{n_l} \sum_{m=n+1}^{N} w_{-n} a_{m-n} \ket{m,B} \nonumber\\
  && \mbox{} + \sum_{m=1}^N w_0^* b_m \ket{m,A}
  + \sum_{n=1}^{n_l} \sum_{m=1}^{N-n} w_{-n}^* b_{m+n} \ket{m,A}
  + \sum_{n=1}^{n_r} \sum_{m=n+1}^{N} w_n^* b_{m-n} \ket{m,A} \nonumber\\
  &=& \epsilon \sum_{m=1}^N a_m\ket{m,A} + \epsilon \sum_{m=1}^N b_m\ket{m,B}.
\end{eqnarray}
This gives $2N$ equations for $2N$ variables $a_m$ and $b_n$, which are given explicitly as
\begin{subequations}\label{eqs}
\begin{eqnarray}
  \label{eq a left}
  \sum_{n=1-m\,(>-n_l)}^{n_r} w_n a_{m+n} &=& \epsilon\, b_m, \qquad
  \text{for}\ m=1,\dots,n_l, \\
  \label{eq a bulk}
  \sum_{n=-n_l}^{n_r} w_n a_{m+n} &=& \epsilon\, b_m, \qquad
  \text{for}\ m=n_l+1,\dots,N-n_r, \\
  \label{eq a right}
  \sum_{n=-n_l}^{N-m\,(<n_r)} w_n a_{m+n} &=& \epsilon\, b_m, \qquad
  \text{for}\ m=N-n_r+1,\dots,N, \\
  \label{eq b left}
  \sum_{n=-n_l}^{m-1\,(<n_r)} w_n^* b_{m-n} &=& \epsilon\, a_m, \qquad
  \text{for}\ m=1,\dots,n_r, \\
  \label{eq b bulk}
  \sum_{n=-n_l}^{n_r} w_n^* b_{m-n} &=& \epsilon\, a_m, \qquad
  \text{for}\ m=n_r+1,\dots,N-n_l, \\
  \label{eq b right}
  \sum_{n=m-N\,(>-n_l)}^{n_r} w_n^* b_{m-n} &=& \epsilon\, a_m, \qquad
  \text{for}\ m=N-n_l+1,\dots,N.
\end{eqnarray}
\end{subequations}
Here, each of \eqnref{eq a bulk} and \eqnref{eq b bulk} gives $N-n_l-n_r$ equations for the lattice points far from the edges; \eqnref{eq a left} gives $n_l$ equations and \eqnref{eq b left} gives $n_r$ equations for the points close to the left edge; \eqnref{eq a right} gives $n_r$ equations and \eqnref{eq b right} gives $n_l$ equations for the points close to the right edge.

Now, let us find the zero-energy ($\epsilon\approx0$) modes. With $\epsilon\approx0$ imposed, $a_m$ and $b_m$ are decoupled in \eqnref{eqs} (this is a consequence of the chiral symmetry). We thus can solve $a_m$ and $b_m$ separately.

To solve the difference equation \eqnref{eq a bulk} for $a_m$, the standard strategy is to make the ansatz $a_m=\xi^m$ with a complex number $\xi$ to be solved. Substituting this ansatz into \eqnref{eq a bulk} with $\epsilon=0$, we have
\begin{equation}\label{solve xi}
\sum_{n=-n_l}^{n_r} w_n \xi^{n} =0,
\end{equation}
which admits those $\xi_j$ in \eqnref{key eq 1} as solutions for $\xi$.
Furthermore, if $\xi_j$ has multiplicity $\nu_j$, any linear superpositions of
\begin{equation}\label{am solutions}
a_m = m^\ell\xi_j^m, \qquad \ell=0,1,\dots,\nu_j-1,
\end{equation}
are also solutions to \eqnref{solve xi}.\footnote{This is because $F(z):=\sum_{n=-n_l}^{n_r} w_n z^{n+n_l}$ can be factorized as $F(z)=(z-\xi_j)^{\nu_j}f(z)$, where $f(z)$ is a polynomial of $z$ and $f(\xi_j)\neq0$, and consequently $\left.\frac{\partial^\ell F(z)}{\partial z^\ell}\right|_{z=\xi_j}= \left.\sum_{n=-n_l}^{n_r} w_n \frac{d^\ell}{dz^\ell}z^{n+n_l}\right|_{z=\xi_j}=0$ for $\ell=1,\dots,\nu_j-1$. This implies that $a_m=\frac{d^\ell}{d\xi_j^\ell}\xi_j^{m+n_l}$ for $\ell=1,\dots,\nu_j-1$ are all solutions to $\sum_{n=-n_l}^{n_r} w_n a_n=0$. These solutions can be reshuffled into \eqnref{am solutions}.}
If $\xi_j=0$, the above solutions all become $a_m=0$, which is problematic and requires closer attention. The fact that $\xi=0$ is a root of $F(z):=\sum_{n=-n_l}^{n_r} w_n z^{n+n_l}$ with multiplicity $\nu_j$ means that $F(z)=z^{\nu_j}f(z)$, where $f(\xi)$ is a polynomial of $\xi$ and $f(0)\neq0$. Consequently, the coefficients of $z^0,z^1,\dots,z^{\nu_j-1}$ all vanish in $F(z)$; in other words, $w_n=0$ for $n=-n_l,-n_l+1,\dots,-n_l+\nu_j-1$.
This implies that \eqnref{eq a bulk} (with $\epsilon=0$) in fact does not involve $a_1,a_2,\dots,a_{\nu_j}$, as the index $m$ in \eqnref{eq a bulk} is delimited by $m\geq n_l+1$.
Therefore, in case of $\xi_j=0$, the variables $a_1,a_2,\dots,a_{\nu_j}$ are completely decoupled from \eqnref{eq a bulk}, and hence there are still $\nu_j$ linearly independent solutions to \eqnref{eq a bulk} given as\footnote{A fully dimerized case with $m>0$ (see \figref{fig:long range amp}) provides a concrete example that $\xi=0$ is a solution of \eqnref{solve xi} and the left edge modes are given by \eqnref{am solutions'}.}
\begin{equation}\label{am solutions'}
a_m = \delta_{m\ell},\quad \ell=1,\dots,\nu_j.
\end{equation}
The candidate solutions as linear superpositions of the form in \eqnref{am solutions} or \eqnref{am solutions'} have to satisfy the boundary conditions \eqnref{eq a left} and \eqnref{eq a right} for the left and right margins.
As there are $\sum_j \nu_j = n_l+n_r$ independently candidate solutions while there are $n_l+n_r$ boundary conditions, we usually do not have a nonzero solution for \emph{exactly} $\epsilon=0$, except for some special conditions (such as a fully demerized limit).\footnote{Even if the boundary conditions coincidentally admit a nonzero solution of $\epsilon=0$, either the solution is not robust or $\epsilon=0$ becomes $\epsilon\approx0$ under small deformations of the hopping amplitudes $w_n$.} Therefore, the zero-energy modes make sense only in the thermal limit $N\rightarrow\infty$.

As $N\rightarrow\infty$, the condition \eqnref{eq a right} demands $a_N\rightarrow0$. Consequently, only the solutions with $\abs{\xi_i}<1$ are valid. Meanwhile, the condition \eqnref{eq a left} gives $n_l$ more equations, which impose further constraints on the coefficients of the linear superposition for the solution. As a result, we have in total $-n_l + \sum_{\abs{\xi_j}<1} \nu_j$ nonzero solutions that are localized at the left edge and exponentially vanish at the right edge, provided $\sum_{\abs{\xi_j}<1} \nu_j \geq n_l$.\footnote{Again, we may accidentally have more nonzero solutions, but only $-n_l + \sum_{\abs{\xi_j}<1} \nu_j$ of them are robust.} Accordingly to \eqnref{key eq 1}, we have just proved that the number of robust zero-energy left edge modes with support in sublattice $A$ is identical to the winding number $\nu$, if $\nu\geq0$.

What if $\nu\leq0$? In this case, \eqnref{eq a left} gives more constraints than the number of the candidate solutions that decay away towards the right edge. Therefore, we have no zero-energy left edge modes with support in sublattice $A$. We should look for the right edge modes instead. Making the ansatz $a_m=\xi^{N-m}$ and substituting it into \eqnref{eq a bulk}, we have
\begin{equation}\label{solve xi inverse}
\sum_{n=-n_l}^{n_r} w_n \xi^{-n} \equiv \sum_{n=-n_r}^{n_l} w_{-n} \xi^n =0,
\end{equation}
which admits those $\xi_i$ in \eqnref{key eq 2} as solutions for $\xi$. Repeating the argument above in the obviously analogous way, we conclude that, according to \eqnref{key eq 2}, the number of robust zero-energy right edge modes with support in sublattice $A$ is identical to $\abs{\nu}$ when $\nu\leq0$.

Similarly, for the zero-energy modes with support in sublattice $B$, there are $\nu$ right edge modes if $\nu\geq0$ and $\abs{\nu}$ left edge modes if $\nu\leq0$, according to \eqnref{key eq 3} and \eqnref{key eq 4}.

When $N$ is finite, the eigenvalue problem \eqnref{eqs} can be solved numerically. The numerical result gives no exactly zero-energy states but only the ``hybridized'' edge states with a small energy splitting around zero, which are with support mostly in sublattice $A$ at the left (right) edge and with support mostly in sublattice $B$ at the right (left) edge. In the thermal limit $N\rightarrow\infty$, the energy splitting vanishes and the hybridized edge states indeed can be reshuffled into ``purified'' edge states with support only in sublattice $A$ or sublattice $B$.

In summary, we have rigorously proved the bulk-boundary correspondence:
\begin{quote}
 In the thermal limit, the winding number $\nu$ of the bulk energy spectrum is identical to the number of robust zero-energy edge modes with support in sublattice $A$ at the left (right) edge or, equivalently, of the robust zero-energy edge modes with support in sublattice $B$ at the right (left) mode, if $\nu\geq0$ ($\nu\leq0$).
\end{quote}

\section{Nonzero-energy edge modes}
The argument above does not exclude the possibility of nonzero-energy edge modes. However, unlike the zero-energy edge modes, the nonzero-energy edge modes, if any, are not robust under adiabatic deformations and therefore are not related to the winding number.

If a system is of the winding number $\nu$, the Hamiltonian $\hat{H}_N$ can always be adiabatically deformed into
\begin{equation}
  \hat{H}_0
  = \sum_{m=1}^{N-\nu} w_\nu^*\ket{m+\nu,A}\bra{m,B}
  + \sum_{m=1}^{N-\nu} w_\nu\ket{m,B}\bra{m+\nu,A}.
\end{equation}
That is, in \eqnref{HN}, all $w_n$ are deformed to zero except that $w_\nu$ is nonzero to have the same winding number $\nu$. This gives a fully dimerized limit (see \figref{fig:long range amp}), for which the energy spectrum is exactly solvable. Obviously, there are $2\nu$ zero-energy modes localized at the left and right edges:
\begin{subequations}
\begin{eqnarray}
\epsilon=0:&&\nonumber\\
\ket{L_m} &:=& \ket{m,A}, \qquad m=1,\dots,\nu,\\
\ket{R_m} &:=& \ket{m,B}, \qquad m=N-\nu+1,\dots,N.\qquad
\end{eqnarray}
\end{subequations}
Meanwhile, we have two nonzero energy eigenvalues $\epsilon=\pm\abs{w_\nu}$, each of which has $(N-\nu)$-fold degenerate eigenstates:
\begin{eqnarray}
&&\epsilon=\pm\abs{w_\nu}:\nonumber\\
&&\quad\ket{\psi^\pm_m} := e^{-i\phi/2}\ket{m+\nu,A}\pm e^{i\phi/2}\ket{m,B},
\qquad m=1,\dots,N-\nu,
\end{eqnarray}
where $w_\nu\equiv\abs{w_\nu}e^{i\phi}$. Those states $\ket{\psi^\pm_m}$ with very small and large $m$ can be viewed as nonzero-energy edge modes.

Now, let us turn on a small perturbation without altering the winding number. Particularly, consider the perturbation Hamiltonian with a small hopping amplitude $w_{\nu'}$ (with $\nu'\neq\nu$ and $\abs{w_{\nu'}}\ll\abs{w_\nu}$):
\begin{equation}
  \hat{H}'
  = \sum_{m=1}^{N-\nu'} w_{\nu'}^*\ket{m+\nu',A}\bra{m,B}
  + \sum_{m=1}^{N-\nu'} w_{\nu'}\ket{m,B}\bra{m+\nu',A}.
\end{equation}
The energy spectrum of $\hat{H}_0+\hat{H}'$ can be approximately solved by the first-order perturbation method. As the eigenstates of $\hat{H}_0$ are degenerate, we have to start with the ``stable'' eigenstates that diagonalize $\hat{H}'$ within the degenerate eigenspace.

It is obvious that $\bra{\psi_1}\hat{H}'\ket{\psi_2}=0$ if $\ket{\psi_{1,2}}$ are any of $\ket{L_m}$ or $\ket{R_m}$. Therefore, $\ket{L_m},\ket{R_m}$ are already the stable eigenstates under $\hat{H}'$. The perturbation theory tells that $\ket{L_m}$ and $\ket{R_m}$ remain the eigenstates of $\hat{H}_0+\hat{H}'$ up to $O(\abs{w_{\nu'}}^2)$ and the first-order energy shift is zero. That is, $\ket{L_m}$ and $\ket{R_m}$ remain to be the zero-energy edge modes.

On the other hand, the nonzero-energy modes $\ket{\psi^\pm_m}$ are not stable under $\hat{H}'$. To find the stable eigenstates that diagonalize $\hat{H}'$ within the $\epsilon=\abs{w_\nu}$ and $\epsilon=-\abs{w_\nu}$ eigenspaces, we have to look for the superposition among the following states:
\begin{equation}
  \dots,\ \ket{\psi^\pm_{m-(\nu-\nu')}},\ \ket{\psi^\pm_m},\ \ket{\psi^\pm_{m+(\nu-\nu')}},\ \ket{\psi^\pm_{m+2(\nu-\nu')}},\ \dots
\end{equation}
The resulting stable states are no longer localized at edge but smeared into bulk. The first-order perturbation under $\hat{H}'$ lifts the degeneracy of $\epsilon=\pm\abs{w_\nu}$ and yields nonzero energy shift. We therefore arrive at the conclusion that nonzero-energy edge modes, if any, are not robust.

\section{Deformations of spatial disorder}
What happens if the system is deformed with small \emph{spatial disorder}? Imposition of spatial disorder cannot be described solely as deformation upon $h(k)$. Rather, it corresponds to replacing the hopping amplitudes $w_n$ with $w_n+\delta w_n(m)$, where $\delta w_n(m)$ are some functions of lattice sites. That is, the total Hamiltonian takes the form
\begin{equation}
  \hat{\mathcal{H}} = \hat{H}_N + \hat{H}' := \hat{H}_N
  + \left.\hat{H}_N\right|_{w_n\rightarrow \delta w_n(m)},
\end{equation}
where $\hat{H}_N$ is given by \eqnref{HN} and $\hat{H}'$ takes the form of $\hat{H}_N$ with $w_n$ replaced by $\delta w_n(m)$. Smallness of $\hat{H}'$ is formally cast as $\delta w := \max_{n,m}\abs{\delta w_n(m)} \ll \Delta E_g$ with $\Delta E_g$ being the bulk spectrum gap.

Within the degenerate zero-energy eigenspace of $\hat{H}_N$, $\hat{H}'$ yields $\bra{\psi_A}\hat{H}'\ket{\psi_A}=\bra{\psi_B}\hat{H}'\ket{\psi_B}=0$ and $\bra{\psi_A}\hat{H}'\ket{\psi_B}\neq 0$, where $\ket{\psi_A}$ ($\ket{\psi_B}$) are zero-energy edge modes with support in sublattice $A$ ($B$). We have shown that $\ket{\psi_A}$ are localized at one edge and exponentially decay towards the other edge, while $\ket{\psi_B}$ behave in the opposite way. Consequently, we have $\bra{\psi_A}\hat{H}'\ket{\psi_B}\sim O(\delta w\, e^{-\lambda N})$, where $\lambda$ is some positive number determined by the decay rates of $\ket{\psi_A}$ and $\ket{\psi_B}$. In the limit $N\rightarrow\infty$, we thus have $\bra{\psi_1}\hat{H}'\ket{\psi_2}\rightarrow0$, where $\ket{\psi_{1,2}}$ are any of $\ket{\psi_A}$ or $\ket{\psi_B}$. The perturbation theory then implies that $\ket{\psi_A}$ and $\ket{\psi_B}$ remain the eigenstates of $\hat{\mathcal{H}}$ up to $O(\delta w^2)$ and the first-order energy shift is zero. That is, the zero-energy edge modes $\ket{\psi_A}$ and $\ket{\psi_B}$ are robust under deformations of spatial disorder provided that the spatial disorder is small enough ($\delta w \ll \Delta E_g$).

\section{Remarks}
So far, we have considered a chain comprised of $N$ sites of type $A$ and $N$ sites of type $B$ as depicted in \figref{fig:SSH model}. Without much difference, our approach can also apply to an ``uneven'' chain comprised of $N+1$ sites of $A$ and $N$ sites of $B$ (i.e., one additional $A$ site is included to the right end in \figref{fig:SSH model}) or the other way around. For an uneven chain, our calculation can be readily repeated, except that \eqnref{eq a right} and \eqnref{eq b right} for the points close to the right edge are slightly modified. In \eqnref{eq a right}, the only change is that the $m=N$ equation is modified from $\sum_{n=-n_l}^0w_na_{N+n}=\epsilon\,b_N$ to $\sum_{n=-n_l}^1w_na_{N+n}=\epsilon\,b_N$;
in \eqnref{eq b right}, the only change is to add one more equation for $m=N+1$ reading as $\sum_{n=1}^{n_r}w^*_nb_{N+1-n}=\epsilon\,a_{N+1}$.
The modified \eqnref{eq a right} still imposes $n_r$ constraints on the linear superposition of candidate edge solutions, while the modified \eqnref{eq b right} now imposes $n_l+1$ constraints.
Consequently, the bulk-boundary correspondence as summarized in the end of \secref{sec:exact calculation} is modified for an uneven chain of $N+1$ sites of $A$ and $N$ sites of $B$ as
\begin{quote}
 In the thermal limit, the number of robust zero-energy edge modes with support in sublattice $A$ ($B$) at the left edge is given by $\abs{\nu}$, and the number of robust zero-energy edge modes with support in sublattice $B$ ($A$) at the right edge is given by $\abs{\nu-1}$, if the winding number $\nu>0$ ($\nu\leq0$).
\end{quote}
Note that there is always at least one edge mode, even if $\nu=0$. This can be viewed as a consequence of the chiral symmetry, which relates an eigenstate with energy $\epsilon$ to a counterpart state with energy $-\epsilon$ and therefore entails the existence of an $\epsilon=0$ state as there are $2N+1$ eigenstates in total.

It should also be remarked that, as commented in the end of \secref{sec:momentum-space Hamiltonian}, to make sense of the smooth approximation of $h(k)$ for a finite system, we have to introduce two positive integers $n_l$ and $n_r$ as the upper bounds for the long-range hopping amplitudes. More precisely, we assume $w_n\approx0$ as long as $n>n_r$ or $n<-n_l$. This is a reasonable assumption, because $w_n$ should become inappreciable when the hopping distance $\abs{n}$ becomes very large. To model a realistic system, $n_l$ and $n_r$ can be chosen in such a way that the condition $\abs{w_{n<-n_l}},\abs{w_{n>n_r}}\ll \frac{1}{n_r+n_l+1}\sum_{-n_l\leq n\leq n_r}\abs{w_n}$ is satisfied. In other words, $n_l$ and $n_r$ provide the cutoffs for safely neglecting far-off hopping.
Also note that the proof of bulk-boundary correspondence relies on the condition $n_l,n_r\ll N$. In case this condition is violated, the conclusion of the bulk-boundary correspondence is no longer valid. As expected, if $N$ is not large enough, some presumed edge modes fail to decay fast enough towards the opposite edge and thus are not counted as localized edge states.
Our investigation just pinpoints how large $N$ must be so that it can be practically viewed as reaching the thermal limit $N\rightarrow\infty$ as the precondition for the bulk-boundary correspondence. In other words, the $N\rightarrow\infty$ limit gives faithful description of edge modes for a finite system, as long as the finite chain length $N$ is much larger than the longest range of hopping, i.e., $n_l,n_r\ll N$.

\begin{acknowledgments}
The authors would like to thank Hsien-Chung Kao for useful discussions. This work was supported in part by the Ministry of Science and Technology, Taiwan under the Grants MOST 104-2112-M-003-012, MOST 105-2811-M-003-028, MOST 106-2112-M-110-010, and MOST 107-2112-M-003-002.
\end{acknowledgments}



\begin{thebibliography}{99}

\bibitem{Hasan:2010xy}
  M.~Z.~Hasan and C.~L.~Kane,
  ``Topological Insulators,''
  Rev.\ Mod.\ Phys.\  {\bf 82}, 3045 (2010)

\bibitem{Qi:2011}
  X.~L.~Qi and S.~C.~Zhang,
  ``Topological insulators and superconductors,''
  Rev.\ Mod.\ Phys.\  {\bf 83}, 1057 (2011)

\bibitem{Laughlin:1981jd}
  R.~B.~Laughlin,
  ``Quantized Hall conductivity in two-dimensions,''
  Phys.\ Rev.\ B {\bf 23}, 5632 (1981).

\bibitem{Hatsugai:1993ywa}
  Y.~Hatsugai,
  ``Chern number and edge states in the integer quantum Hall effect,''
  Phys.\ Rev.\ Lett.\  {\bf 71}, no. 22, 3697 (1993).

\bibitem{Hatsugai:2009ywa}
  Y.~Hatsugai,
  ``Bulk-edge correspondence in graphene with/without magnetic field: Chiral symmetry, Dirac fermions and edge states,''
  Solid.\ State.\ Comm.\ {\bf 149}, 1061 (2009).

\bibitem{Essin:2011ywa}
  A.~M.~Essin and V.~Gurarie,
  ``Bulk-boundary correspondence of topological insulators from their Green’s functions,''
  Phys.\ Rev.\ B.\  {\bf 84}, no. 12, 125132 (2011).

\bibitem{Graf:2013ywa}
  G.~M.~Graf and M.~Porta,
  ``Bulk-edge correspondence for two-dimensional topological insulators,''
  Commun.\ Math.\ Phys.\  {\bf 324}, no. 3, 851 (2013).

\bibitem{Rudner:2013ywa}
  M.~S.~Rudner, N.~H.~Lindner, E.~Berg, and M.~Levin,
  ``Anomalous Edge States and the Bulk-Edge Correspondence for Periodically Driven Two-Dimensional Systems,''
  Phys.\ Rev.\ X.\  {\bf 3}, no. 3, 031005 (2013).

\bibitem{Cano:2014ywa}
  J.~Cano, M.~Cheng, M.~Mulligan, C.~Nayak, E.~Plamadeala, and J.~Yard,
  ``Bulk-edge correspondence in (2+1)-dimensional Abelian topological phases,''
  Phys.\ Rev.\ B.\  {\bf 89}, no. 11, 115116 (2014).

\bibitem{Prodan:book}
 E.~Prodan and H.~Schulz-Baldes,
 {\it Bulk and Boundary Invariants for Complex Topological Insulators: From $K$-Theory to Physics}, (Springer, Switzerland 2016).

\bibitem{Su:1979}
  W.~P.~Su, J.~R.~Schrieffer, A.~J.~Heeger,
  ``Solitons in Polyacetylene,''
  Phys.\ Rev.\ Lett.\ {\bf 42}, 1698 (1979).

\bibitem{Heeger:1988zz}
  A.~J.~Heeger, S.~Kivelson, J.~R.~Schrieffer and W.-P.~Su,
  ``Solitons in conducting polymers,''
  Rev.\ Mod.\ Phys.\  {\bf 60}, 781 (1988).

\bibitem{Jackiw:1976}
  R.~Jackiw and C.~Rebbi,
  ``Solitons with fermion number $1/2$,''
  Phys.\ Rev.\ D {\bf 13}, 3398 (1976).

\bibitem{Ryu:2010zza}
  S.~Ryu, A.~P.~Schnyder, A.~Furusaki and A.~W.~W.~Ludwig,
  ``Topological insulators and superconductors: Tenfold way and dimensional hierarchy,''
  New J.\ Phys.\  {\bf 12}, 065010 (2010).

\bibitem{Asboth:book}
 J.~K.~Asb\'{o}th, L.~Oroszl\'{a}ny, A.~P\'{a}lyi,
 {\it A Short Course on Topological Insulators: Band-structure topology and edge states in one and two dimensions}, (Springer, Switzerland 2016).

\bibitem{Pershoguba:2012}
 S.~S.~Pershoguba and V.~M.~Yakovenko,
 ``Shockley model description of surface states in topological insulators,''
 Phys.\ Rev.\ B {\bf 86}, 075304 (2012).

\bibitem{Rhim:2017}
 J.~W.~Rhim, J.~Behrends, J.~H.~Bardarson,
 ``Bulk-boundary correspondence from the intercellular Zak phase,''
 Phys.\ Rev.\ {bf B} 95, 035421 (2017).

\bibitem{Rudin:1976}
 W.~Rudin, {\it Principles of Mathematical Analysis},
 (McGraw-Hill, New York 1976).


\end{thebibliography}
\end{document}